\documentclass[12pt]{article}
\begin{document}
\begin{flushright}
MRI-PHY/P980553
\end{flushright}
\begin{center}
\vspace*{1.0cm}

{\LARGE{\bf  $S$-matrices and bi-linear sum rules of}}
\vskip 0.2cm
{\LARGE{\bf conserved charges in affine Toda field theories}}

\vskip 1.5cm

{\large {\bf S. Pratik  Khastgir\footnote{email: pratik@mri.ernet.in}}}

\vskip 0.5cm

{\sl Mehta Research Institute for Mathematics and Mathematical Physics,} \\
{\sl Chhatnag Road, Jhusi, Allahabad 211 019, INDIA.}

\end{center}

\vspace{1 cm}

\begin{abstract}
\noindent 
The exact quantum $S$-matrices and conserved charges are known for 
affine Toda field theories(ATFTs). In this note we report on a new
type of bi-linear sum rules of conserved quantities derived from these
exact $S$ matrices. They exist when there
is a multiplicative identity among $S$-matrices of a particular ATFT.
Our results are valid for simply laced as
well as non-simply laced ATFTs. We also present a few explicit examples. 


\end{abstract}

\vspace{1 cm}
\def\ts{\thinspace}
\newcommand\et{{\sl et al. }}
\newcommand{\mbold}{\mbox{\boldmath$m$}}
\newcommand\beq{\begin{equation}}
\newcommand\eeq{\end{equation}}
\newcommand\Bear{\begin{eqnarray}}
\newcommand\Enar{\end{eqnarray}}
\newcommand{\rref}[1]{(\ref{#1})}
\def\hcd#1{\{ #1 \}'{}}
\newcommand\Zam{Zamolodchikov}
\newcommand \ZZ {A. B. \Zam\  and Al. B. \Zam, {\it Ann. Phys.}
{\bf 120} (1979)  253}
\newcommand\CMP{{\it Comm.\ts Math.\ts Phys.\ts}}
\newcommand\IJMP{{\it Int.\ts J.\ts Mod.\ts Phys.\ts}}
\newcommand\NP{{\it Nucl.\ts Phys.\ts}}
\newcommand\PL{{\it Phys.\ts Lett.\ts}}
\newcommand\Zm{Zamolodchikov}
\newcommand\AZm{A.\ts B.\ts \Zm}
\newcommand\dur{H.\ts W.\ts Braden, E.\ts Corrigan, P.\ts E.\ts Dorey 
and R.\ts Sasaki}


Affine Toda field theory\footnote{For an excellent review see
Ref. \cite{Co} }\cite{MOPa} is a massive scalar field theory with 
exponential interactions in $1+1$ dimensions described by the Lagrangian
\begin{equation}
{\cal L}={1\over 2}
\partial_\mu\phi^a\partial^\mu\phi^a-{m^2\over
\beta^2}\sum_{i=0}^rn_ie^{\beta\alpha_i\cdot\phi}.
\label{ltoda}
\end{equation}
The field $\phi$ is an $r$-component scalar field, $r$ is the rank of a
compact semi-simple Lie algebra $G$ with $\alpha_i$;
$i=1,\ldots,r$ being its simple roots and $\alpha_0$ is the affine root. 
The Kac-Coxeter labels $n_i$ are such that $\sum_{i=0}^rn_i\alpha_i=0$, 
with the convention $n_0=1$. The quantity, $\sum_{i=0}^rn_i$, 
is denoted by `$h$' and known as the Coxeter number.
`$m$' is a real parameter setting the mass scale of the theory 
and $\beta$ is a real coupling constant,
which is relevant only in quantum theory.

Integrable field theories(such as ATFTs) are characterised by an 
infinite set 
of conserved quantities. In ATFTs, it is well-known that these 
conserved quantities are related with 
the Cartan matrix of the associated finite Lie algebra 
(see eqn. \rref{fp}).
In this note we report that in certain circumstances these
conserved quantities satisfy interesting bi-linear sum rules,
which we believe have never been encountered in theoretical physics.

Quantum $S$-matrices for all simply laced [3--8] as well as non-simply
laced \cite{DGZc} affine Toda theories are known.
 Based on the assumption that the infinite set of conserved quantities
be preserved after quantisation, only the elastic processes are
allowed and the multi-particle $S$-matrices are factorised into 
a product of two particle elastic $S$-matrices.
A typical elastic, unitary $S$-matrix for a process $a+b\rightarrow a+b$
can be written as product of ratios of hyperbolic sines.
\beq
S_{ab}(\theta)=\prod_{x\in I_{ab}} \{x\}, \qquad 
\{x\}={{(x-1)(x+1)}\over{(x-1+B)(x+1-B)}},
\label{smat}
\eeq         
for some set of integers $I_{ab}$. The building block $(x)$ and 
the function 
$B(\beta)$ are given by
\beq
(x)={{\sinh({\frac{\theta}{2}+\frac{i\pi}{2h}x)}}\over,~~{\rm etc.}
{\sinh({\frac{\theta}{2}-\frac{i\pi}{2h}x)}}},\qquad
B(\beta)={{\frac{1}{2\pi}}\frac{\beta^2}{1+\beta^2/4\pi}}.
\label{block}
\eeq
$\theta=\theta_a-\theta_b$ is the relative rapidity
($p_a\equiv(m_a\cosh{\theta_a},m_a\sinh{\theta_a})$), $h$ is the Coxeter 
number of the Lie algebra on which theory is based.
The above $S$-matrices respect crossing symmetry
and bootstrap principle \cite{BCDSc}.
Bootstrap equations will constrain both $S$-matrix elements, 
$S_{ab}(\theta)$, as well as
eigenvalues, $q_s^a$, of the conserved charges acting on single particle
states (defined by $Q_s|A_a(\theta)\rangle=q_s^ae^{s\theta} 
|A_a(\theta)\rangle$, with
 $A_a(\theta)$ denoting a particle type $a$). For the fusion process
$a+b\rightarrow \bar c$, bootstrap equation for the $S$-matrices is the
following \cite{Zam},
\beq 
S_{d\bar c}(\theta)=S_{da}(\theta-i{\bar \theta}_{ac}^b)
S_{db}(\theta+i{\bar \theta}_{bc}^a),\label{boots}
\eeq  
where ${\bar \theta}$s are certain angles defined for every triplet of
particles possessing a non vanishing three-point coupling $C^{abc}$.
For the conserved charges it has the following form (is in fact
satisfied by the logarithmic derivative of the $S$-matrix \cite{BCDSc}),
\beq
q_s^{\bar c}=q_s^ae^{-is{\bar \theta}_{ac}^b}+q_s^b
e^{is{\bar \theta}_{bc}^a}.
\label{qboot}\eeq
$q_s^{\bar a}=(-1)^{s+1}q_s^a$
is the effect of charge conjugation on the conserved charges.
Nontrivial solutions to the conserved charge bootstrap only occur if
the spin $s$ modulo $h$, is equal to an exponent of rank-$r$ group
$G$ on which ATFT is based. Furthermore each of the $r$ particles 
of the simply laced theories is associated unambiguously with the 
spots on the unextended Dynkin diagram of $G$ and thus to the simple
roots($\alpha_i$) of the associated finite Lie algebra.  
Association happens in such a way that a vector made out of the
conserved charges with spin $s$, i.e. ${\bf {\bar
q}_s}=(q_s^{\alpha_1},q_s^{\alpha_2},...,q_s^{\alpha_r})$,
forms an eigenvector of the Cartan matrix of $G$, with eigenvalue
$2-2\cos({\pi s/h})$ \cite{KM}. Thus,
\beq 
C{\bf {\bar q}_s}=\lambda_s{\bf {\bar q}_s},\qquad \lambda_s 
=2-2\cos({\pi s/h})
\label{fp}
\eeq
where $C$ is the Cartan matrix
$C_{ij}=2\alpha_i\cdot\alpha_j/\alpha_j\sp2$,  
$i,j=1,\dots ,r. $
The complete list of eigenvectors for the various simply laced
theories can be found in the Table-2 of Ref. \cite{Dorey}. The spin-1
charge vector gives the masses of the various
particles(i.e. $q_1^a=m_a$).
For non-simply laced theories these charges were calculated in the Ref.
\cite{CDS}.

Before stating the results we would consider two examples. The
following one is from $e^{(1)}_{6}$ theory.
The vector ${\bf {\bar q}_s}$ is given by 
(see Table 2 of Ref.\cite{Dorey}.) 
\beq
{\bf {\bar q}_s}=\pmatrix{q^l_s\cr
                          q^{\bar h}_s\cr
                          q^L_s\cr
                          q^H_s\cr
                          q^h_s\cr
                          q^{\bar l}_s\cr}=\pmatrix{\sin 11\theta_s\cr
                                                    \sin 10\theta_s\cr
                                   \sin 8\theta_s-\sin 2\theta_s\cr
                                                     \sin 3\theta_s\cr
                                                      \sin 2\theta_s\cr
                                                      \sin
                                                      \theta_s\cr}\qquad
  {\rm with}~~ \theta_s =(\pi s/12).
\eeq

The following identity exists in $e^{(1)}_{6}$\footnote{see Table 1 of
  Ref. \cite{BCDSc}}:
\begin{equation}
S_{Hh}(\theta)=S_{hl}(\theta)S_{{\bar h}l}(\theta)S_{Ll}(\theta)
\end{equation}
and we observe that,
\begin{equation}
q^H_sq^h_s=q^{h}_sq^{l}_s+q^{{\bar h}}_sq^{l}_s+q^{L}_sq^{l}_s
\end{equation}
for $s=1,4,5,7,8,11$.
For $s=1$, above equation becomes,
\begin{equation}
m_Hm_h=m_{h}m_{l}+m_{{\bar h}}m_{l}+m_{L}m_{l}.
\end{equation}

Next we choose an example from non-simply laced theories.
We consider the dual pair $(f^{(1)}_4,~e_6^{(2)})$, Ref. \cite{CDS},
\beq
{\bf {\bar q}_s}=\pmatrix{q^1_s\cr
                          q^2_s\cr
                          q^3_s\cr
                          q^4_s\cr}=
            \pmatrix{\sin ({{s\pi}\over H})\sin({{2s\pi}\over H'})\cr
                          \sin({{3s\pi}\over H})\sin({{s\pi}\over H'})\cr
                         \sin({{2s\pi}\over H})\sin({{2s\pi}\over H'})\cr
            \sin ({{3s\pi}\over H})\sin({{2s\pi}\over H'})\cr },~~
  {\rm where}~~{1\over H}+{1\over H'} ={1\over 6}~~{\rm and}~H=12+3B.
\eeq
In this case we have the following identity\footnote{A list of few
  multiplicative identities of the $S$-matrix for different ATFTs is
  given in appendix A}
\beq
       S_{33}(\theta)= S_{11}(\theta) S_{14}(\theta).
\eeq
One can again observe,
\beq
(q^3_s)^2=(q^1_s)^2+q^1_sq^4_s \qquad{\rm for}~~s=1,5~{\rm mod}~6.
\eeq
The bi-linear mass relation in this case reads,
\beq
m^2_3=m^2_1+m_1m_4
\eeq
In the above equation masses should be understood  as the floating
masses (defined in expression (4.1) of Ref. \cite{CDS}) for the common
quantum theory of the dual pair  $(f^{(1)}_4,~e_6^{(2)})$.

Observing these examples carefully we make the following proposition. To
prove it we would need to make certain conjecture about the Fourier
coefficients in the expansion of logarithmic derivatives of $S$-matrix.

{\bf{Proposition}:}
              
If
\begin{equation}
\prod_{a,b\in\{i,j\}}S_{ab}(\theta)=\prod_{a',b'\in \{i',j'\}} 
S_{a'b'}(\theta)
\label{imult}\end{equation}

for some sets $\{i,j\}$ and $\{i',j'\}$ then,
\begin{equation}
\sum_{a,b\in \{i,j\}}q_s^aq_s^b=\sum_{a',b'\in \{i',j'\}}q^{a'}_sq^{b'}_s
\label{iadd}\end{equation}
 
for every individual $s$.

{\bf{Proof}:} Starting with the logarithmic derivative of the
$S$-matrix (Ref. \cite{BCDSc}), $T_{ab}\equiv{d\over{d\theta}}\ln
S_{ab}(\theta) $, one can write eqn. \rref{boots} in terms of the
quantities $T$ to obtain,
\beq 
T_{d\bar c}(\theta)=T_{da}(\theta-i{\bar \theta}_{ac}^b)+
T_{db}(\theta+i{\bar \theta}_{bc}^a)\label{tboots}.
\eeq  
Expanding $T(\theta)$ as a Fourier series in $\theta$($S(\theta)$ is
$2\pi i$ periodic),
\beq
T_{ab}(\theta)=\sum_{s=-\infty}^{\infty}t_s^{ab}e^{s\theta},
\eeq
where quantities $t_s^{ab}$ are a set of coefficients which as a
consequence of \rref{tboots} must satisfy
\beq
t_s^{d\bar c}=t_s^{da}e^{-is{\bar \theta}_{ac}^b}+t_s^{db}
e^{is{\bar \theta}_{bc}^a}
\label{tqboot},
\eeq
very reminiscent of the eqn. \rref{qboot}. The unitarity and crossing
symmetry conditions put the following restrictions on the coefficients
 $t_s^{ab}$. 
\beq
T_{ab}(\theta)=T_{ab}(-\theta) ~~\Rightarrow~~
 t_s^{ab}=t_{-s}^{ab}, \label{ucond}\eeq
and 
\beq
T_{ab}(i\pi-\theta)=-T_{b\bar a}(\theta) ~~\Rightarrow~~
 t_s^{ab}=(-1)^{s+1}t_{-s}^{b\bar a}. \label{ccond}\eeq

From the form of eqn. \rref{tqboot} we are led to the following
factorised form of the Fourier coefficients\footnote{There may 
be some multiplicative coefficient
independent of $a,b$ and $s$ on the right hand side of the expression 
\rref{conjec}}, 
\beq
t_s^{ab}\equiv q_s^a q_s^b\label{conjec}
\eeq 
which is consistent with all the above constraints.
The eqn. \rref{tqboot} will immediately follow from the 
eqn. \rref{qboot}.

Now taking logarithmic derivative of both sides of the equation
\rref{imult} we have,
\Bear 
\sum_{{a,b}\in\{i,j\}}T_{ab}(\theta) &=&\sum_{{a',b'}\in\{i',j'\}}
T_{a'b'}(\theta)\nonumber\\
\sum_{{a,b}\in\{i,j\}}\sum_{s=-\infty}^{\infty}t_s^{ab}e^{s\theta} &=
&\sum_{{a',b'}\in\{i',j'\}}\sum_{s=-\infty}^{\infty}
t_s^{a'b'}e^{s\theta}
\nonumber\\
\sum_{s=-\infty}^{\infty}(\sum_{{a,b}\in\{i,j\}}t_s^{ab})e^{s\theta} &=
&\sum_{s=-\infty}^{\infty}(\sum_{{a',b'}\in\{i',j'\}}
t_s^{a'b'})e^{s\theta}
\nonumber\\
\sum_{{a,b}\in\{i,j\}}t_s^{ab} &=
&\sum_{{a',b'}\in\{i',j'\}}t_s^{a'b'}.
\label{proof}
\Enar
In the last equation we use \rref{conjec} to arrive at the result
\rref{iadd}.

To conclude we have shown an interesting relation between the
S-matrices and the conserved charges. One may derive these results
from the work of Niedermaier (Ref. \cite{Nied}). But there author
talks about the exact quantum charges of simply laced theories. 
Our results are valid for non-simply laced theories as well.
Moreover we think that these bi-linear sum rules are not something
sacred to the real coupling ATFTs, but would also work for other
two-dimensional integrable models like sine-Gordon theory
(which is actually an imaginary coupling $a_1^{(1)}$ ATFT), massive 
Thirring model etc.. One of the main features of the these models
is topological solitons. So, analogously we would like to think that 
the bi-linear sum rules would connect topological charges of the
solitons when there are multiplicative identities present in 
soliton-soliton scattering matrix.

\section*{Acknowledgements} 
 I would like to thank Prof. Ryu Sasaki for reading  the manuscript 
carefully and for making valuable comments and
suggestions at various stages of the work.

\section* {Appendix A:}
List of few multiplicative identities of the
$S$-matrix for different ATFTs.

\setcounter{equation}{0}
\renewcommand{\theequation}{A.\arabic{equation}}
\vspace{1 cm}
i) $a_r^{(1)}, d_r^{(1)}, (c_r^{(1)}, d_{r+1}^{(2)})~ {\rm and }~
(b_r^{(1)}, a_{2r-1}^{(2)}):$

\Bear
&S_{22}(\theta)&=S_{11}(\theta)S_{13}(\theta)\\
&S_{23}(\theta)&=S_{12}(\theta)S_{14}(\theta)\\
&S_{24}(\theta)&=S_{13}(\theta)S_{15}(\theta\\
&S_{33}(\theta)&=S_{22}(\theta)S_{15}(\theta)=S_{11}(\theta)
S_{13}(\theta)S_{15}(\theta)\\
&S_{34}(\theta)&=S_{23}(\theta)S_{16}(\theta)=S_{12}(\theta)
S_{14}(\theta)S_{16}(\theta),~~{\rm etc.}
\Enar

ii) $e_6^{(1)}:$

\Bear
&S_{LL}(\theta)&=S_{ll}(\theta)S_{l\bar l}(\theta)\\
&S_{HL}(\theta)&=S_{hl}(\theta)S_{{\bar h} l}(\theta)
=S_{h\bar l}(\theta)
S_{{\bar h}\bar l}(\theta)=S_{hl}(\theta)S_{h{\bar l}}(\theta)
=S_{{\bar h}l}(\theta)S_{{\bar h}\bar l}(\theta)\\
&S_{HH}(\theta)&=S_{hh}(\theta)S_{h\bar h}(\theta)
=S_{{\bar h}\bar h}(\theta)
S_{h\bar h }(\theta),~~{\rm etc.}
\Enar

iii) $e_7^{(1)}:$

\Bear
&S_{33}(\theta)&=S_{22}(\theta)S_{23}(\theta)\\
&S_{54}(\theta)&=S_{12}(\theta)S_{34}(\theta)\\
&S_{46}(\theta)&=S_{14}(\theta)S_{12}(\theta)S_{23}(\theta)\\
&S_{11}(\theta)S_{77}(\theta)&=S_{55}(\theta)
S_{66}(\theta),~~{\rm etc.}
\Enar
 
iv) $e_8^{(1)}:$

\Bear
&S_{23}(\theta)&=S_{16}(\theta)\\
&S_{22}(\theta)&=S_{11}(\theta)S_{12}(\theta)\\
&S_{45}(\theta)&=S_{17}(\theta)S_{23}(\theta)\\
&S_{67}(\theta)&=S_{15}(\theta)S_{25}(\theta)
S_{34}(\theta),~~{\rm etc.}
\Enar


\begin{thebibliography}{99}
\bibitem{MOPa}{A. V. Mikhailov, M. A. Olshanetsky and A. M. Perelomov},
{\CMP\ts  {\bf 79} (1981) 473}; \\
G.\ts Wilson,  
{\it Ergod. Th. and Dynam. Sys.} {\bf 1} (1981) 361;\\
{D.\ts I.\ts Olive and N.\ts Turok}, {\NP\ts {\bf B265} (1986) 469}.
\bibitem{Co} {E. Corrigan, {\it Recent developments  in affine Toda
quantum field theory}, Invited lecture at the CRM--CAP summer school
`Particles and Fields 94', Banff, Alberta, Canada, Durham preprint,
DTP-94/55, hepth/9412213.}
\bibitem{AFZa}A.\ts E.\ts Arinshtein, V.\ts A.\ts Fateev and \AZm,
{\PL\ts {\bf B87} (1979) 389.}
\bibitem{BCDSa} \dur, \PL\ts {\bf B227} (1989) 411.
\bibitem{BCDSc} \dur, \NP {\bf B338} (1990) 689; \\
H.\ts W.\ts Braden and R.\ts Sasaki, \PL {\bf B255} (1991) 343; \\
R.\ts Sasaki and F.\ts P.\ts Zen, \IJMP {\bf 8} (1993) 115.
\bibitem{CMa}{P.\ts Christe and G.\ts Mussardo},
\NP\ts {\bf B330} (1990) 465; 
\IJMP {\bf A5} (1990) 4581.
\bibitem{DDa}{C.\ts Destri and H.\ts J.\ts de Vega},
 {\it Phys. Lett.} {\bf B233} (1989) 336.
\bibitem{KM} {T.R. Klassen and E. Melzer, {\it Nucl. Phys.} {\bf B338}
(1990) 485.}
\bibitem{DGZc}{G.\ts W.\ts Delius, M.\ts T.\ts Grisaru and D.\ts Zanon},
\NP\ts {\bf B382} {(1992) 365}.
\bibitem{Zam}\AZm,\IJMP {\bf 4} (1989) 4235.
\bibitem{Dorey}{P. Dorey, {\it Nucl. Phys.} {\bf B358}
(1991) 654.} 
\bibitem{CDS}{E.\ts Corrigan, P.\ts E.\ts Dorey and R.\ts Sasaki},
 {\NP\ts {\bf B408} (1993) 579.}
\bibitem{Nied}{M. R. Niedermaier, {\it Nucl. Phys.} {\bf B424}
(1994) 184.} 

\end{thebibliography}
\end{document}